**Why is neural connection weight a weak predictor of correlated neural activity?**

Daniel Graham[1]

[1] Department of Psychological Science, Hobart & William Smith Colleges, Geneva, NY, 14456 USA; graham@hws.edu

**ABSTRACT:**
**As the field of connectomics has matured, it has expanded from mapping the existence of connections between brain components to measuring the strength of connections. This information is increasingly accessible via methodologies such as pairing functional magnetic resonance (MR) imaging and MR tractography in the same human subject, as well as novel methods in non-human animals using optogenetics. Systems and network neuroscience have in recent years focused extensively on explaining correlation patterns of functional activity in the brain in terms of the degree of connectedness of brain components, the so-called functional connectivity-structural connectivity relationship (SC-FC). What has been surprising has been how low the SC-FC correlations are. Why is it that brain parts that are more well-connected appear not to engender more correlated activity between them? Several explanations have been proposed but one possibility has not been considered: perhaps more neural activity doesn't imply more functional involvement. This article examines this possibility and proposes a new general framework for understanding brain network dynamics based on the design constraints of large-scale network communication systems. With this new perspective, we may start to answer the article's main question, and perhaps others.**

As the field of connectomics has matured, it has expanded from mapping the existence of connections between brain components to measuring the strength of connections. This information is increasingly accessible via methodologies such as pairing functional magnetic resonance (MR) imaging and MR tractography in the same human subject, as well as novel methods in non-human animals using optogenetics. Systems and network neuroscience have in recent years focused extensively on explaining correlation patterns of functional activity in the brain in terms of the degree of connectedness of brain components, the so-called functional connectivity-structural connectivity relationship (SC-FC). The standard—but often unspoken—rationale for this approach is that areas that show strong associations in functional activity need a high volume of linkages in order to deliver a high number of signals between the two areas during coordinated activation.

What has been surprising has been how low SC-FC correlations tend to be: in current state-of-the-art tests, the strength of connection predicts at best 20% of the variance in activity for regions connected by white matter tracts (human MR imaging; Seguin et al., 2020) and less than 3% at the finer-grained synaptic connection level (using transgenic *C. elegans*, Yemini et al., Cell, 2021). Many manipulations such as looking at only stimulus-related (or non-stimulus-related) activity, ignoring the weakest connections, examining activity rankings instead of magnitudes, and scaling data with nonlinear transforms, have been tried, but seem to provide little improvement (Yemini et al., 2021). Why is this the case?

Even without appealing to mechanistic explanations, we can partially explain the low association in terms of methodology. As is well known, MR imaging has strong limitations in resolution, which limit both functional and structural measures derived from MR signals. For example, spatial resolution is such that activation signals summarize hundreds of thousands or millions of neurons. Temporal limitations preclude knowledge of which of two areas became active first, or if they were active simultaneously. As a result, researchers typically measure zero-lag correlations (e.g., Vázquez-Rodriguez et al., 2019). This lack of precision in measuring functional activity is compounded by strong limitations in structural imaging, which generate substantial false positives (Maier-Hein et al., 2017) and preclude measurement of the directionality of connections (see Kale et al., 2020). Invasive structural and functional measures in non-human animals are promising but also have significant sampling and resolution limitations (Callaway and Garg, 2017). Thus, part of the problem may be that current measures are not powerful enough to find an assumed correlation in magnitude between structure and functional activity.

A related problem comes with respect to the definition of functional connectivity: does it require co-activation or sequential activation? From a dynamical systems perspective, co-activation and sequential behavior are distinct processes. Co-activation generally implies synchrony and entrainment and/or prior activation of a common input (Messé et al., 2015; Garcia et al., 2012). In contrast, sequential activation tends to imply finely-tuned time delays and recurrent pathways. So until there is agreement about precisely what is meant by "functional connectivity" between units, any degree of correlation between functional and structural connectivity will remain ambiguous.

But let's assume it were possible to determine when two parts of the brain show a genuine relationship between precise, well-defined measures of coordinated functional activity and accurate, directed measures of connection strength. I predict that, even if this were achievable, we would still observe low SC-FC correlations.

In part, this may be due to the complexity of vertebrate intrabrain signaling. Neural activity in a given population may depend to a greater extent than is usually recognized on non-classical signaling mechanisms, as has been shown in invertebrates (Bargmann and Marder, 2013). The path of these signals would not necessarily correspond to chemical and electrical synapses; such signals can diffuse over a significant area and be delivered by multiple means (e.g., neuroendocrine cells and hormones). In addition, in situations where neurotransmitters and their receptors can be accurately identified in single synapses, there is a surprisingly high occurrence of what we might call senders without receivers and receivers without senders (Yemini, 2021). Where this occurs, it is presumed that sender and receiver are not synaptic partners but instead pass signals to more distant neurons.

Other fundamental neurobiological signaling mechanisms may play a role in generating low SC-FC correlations. A given population may need inputs from multiple other populations simultaneously, not just from the population it is best connected to. However, if this is the case, we should still see reasonably strong correlations in functional activity between the best-connected areas. Low SC-FC correlation can't simply be a result of a global need for multiple inputs to initiate high activity in a given area.

But one explanation for a fundamentally low SC-FC relationship has not been considered: perhaps more activity in a neural population doesn't necessarily imply more involvement in a functional task. If more activity in neural populations is not necessarily associated with more effect on functional outcomes, there would be no need to deliver such increases in activity all at once over many parallel linkages. Dense connectivity would in this case need a different explanation.

The novelty of this view is probably due to a long-standing theoretical bias in the field: brains are assumed to compute things (von Neumann, 1958). In this view, the brain is more active when it is doing more computing, be it for the purpose of perceiving an object, planning a motor action, or "retrieving" a memory. Contemporary deep learning theories also assume this: for example, convolutional neural network models (e.g., Yamins et al., 2014) aim to predict when visual neurons have higher spike rates based on when a computational model trained for a similar task generates higher numerical outputs.

Thus, in computation, more usually implies more. More activity—and therefore more energy—are needed each time we want to do something. However, if coordinated activity corresponds to passing the results of computations about whether a partner neural population should also be active, the brain's sharply limited energy budget would make this a major challenge. The brain can't have all or even most of the sending units in a given area simultaneously active on a regular basis. The brain would have to expend a great deal of energy all at once across a large population. However, we know, given the overwhelming sparseness of activity across populations, that it does not operate this way (Levy and Baxter, 1996; Lennie, 1998; Attwell and Laughlin, 2001; Field, 1994; Olshausen and Field, 2005; Graham and Field, 2006; for recent experimental confirmation, see the study of activity in 60,000 visual neurons in the mouse by de Vries et al., 2020). This is just as much the case if neural populations require multiple simultaneous inputs or just one input. Thus, parallel linkages may not be built to deliver a barrage of activity across channels all at once.

Perhaps the *quantity* of signals—as opposed to their presence or timing—actually tells us little. Analogizing brain signaling with a communication system like the internet helps illustrate an alternative approach. Say we were to intercept and decode a message from the internet. The message could say "I am on vacation and will return next week" or it could say "A bomb will explode in the stadium at noon." The two messages obviously lead to starkly different outcomes: the former implies do nothing, while the latter would evoke a massive mobilization of resources. Yet the two messages are the same magnitude: they contain the same number of bits.

We don't typically think of brains in these terms because we see things from the point of view of computations, rather than from the point of view of messages. The former occur in sequence (or in parallel sequences), each operating on the outcome of the last, and often in a hierarchy of ascending complexity. Messages traveling on large, well-connected networks, on the other hand, pass from node to node carrying essentially the same information, though they may change their material form (graded potentials, varieties of spike codes, chemical signals, etc.). A message's essential characteristics are that it can be generated at will and delivered reliably to a remote receiver elsewhere on the network. The sheer magnitude of a message means little: unlike in computing, in communication more does not necessarily mean more.

Thinking about the brain as an internet-like system for exchanging messages has the benefit of emphasizing the need for flexibility, which is another attribute lacking in prototypical computers (Poggio, 1984). Though absent from classical theory based on McCulloch Pitts-like computational models of neurons, flexibility is increasingly seen as a core brain function that depends on task demands and is accomplished through multiple means. As first postulated by Waxman (1972) and Scott (1977), neurons have many mechanisms that can potentially route signals in a selective manner over time (e.g., Gollisch and Meister, 2010; Sardi et al., 2017; Gidon et al., 2020). The same is true of larger populations (e.g., Cole et al., 2013, Gerraty et al., 2018; Palmigiano et al., 2017). Ephaptic interactions among white matter axons (Sheheitli and Jirsa, 2020) are another possible mechanism for selective routing. Given the necessity of this kind of flexibility, and the changing patterns of coordinated activity that flexibility engenders, it does not make sense to expect activity patterns to systematically match a pattern of static, direct connections (Chiêm et al., 2020).

Moreover, messages need not communicate information directly related to any upcoming functional outcome (e.g., "muscleless synergies" in motor systems; Mohan et al., 2019). Again, the internet provides a useful analogy. Routers on the internet constantly exchange a variety of signals whose purpose is to ensure efficient message-passing across the entire network. These include acknowledgments (ACKs), which relay message receipt back to the sender; keep-alives, which indicate to neighbors that a given router is functioning normally; and pings (echo requests), which probe paths to more distant nodes. Such messages are small and brief, and carry no message "content," but are crucial for system function. If we looked only for signals with the greatest magnitude in an internet-like system, these system-maintenance signals would be missed—and the system as a whole misunderstood (Graham, 2021).

We shouldn't expect the brain to function in exactly the same way as the internet. But we may see that the internet's tricks are used to address similar challenges in the brain. Though the space of possible routing strategies is large, the internet provides a model of an extraordinarily successful and efficient protocol for flexible, selective mass intercommunication. Indeed, a network of billions of unit like our brain, where each unit is just a few linkages away from practically any other, and where flexibility of function from moment to moment is paramount, has an obvious need to solve internet-like communication problems. It is too early to specify the neural mechanisms that address these challenges, though as I propose, looping thalamocortical architecture is a good place to look, especially for ACK-like message-receipt signals (Graham and Rockmore, 2011; Graham, 2014; Graham, 2021).

From the viewpoint of the internet metaphor, then, high connectivity does not exist to provide lots of synchronized input, leading to correlated activity. Instead, it is a form of multiplexing in space and time: it allows many different messages to be sent selectively over time to different destinations, rather than all at once. We should assume that, like the internet, most channels in the brain operate far below capacity the vast majority of the time. This means activity in one area will not necessarily lead to time-adjacent activity in areas that appear to be their primary target based on connection strength. The point is that we would not think to look for such behavior without a new metaphor that goes beyond the computer.

Seguin et al. (2020) have already shown that communication models are a fruitful approach (see also Mišić et al., 2014; Hao and Graham, 2020; Graham et al., 2020). Seguin et al. (2020) found that models of connectivity that include simple communication protocols such as seeking locally short paths and taking random walks on the network substantially outperformed the best models that posit no overarching communication strategy in predicting co-activation. While this study used simple communication models that were not explicitly inspired by the internet, it did use the strategy of intuiting a communication protocol with particular goals and constraints to understand whole-brain activity. It is a step in the right direction.